\documentclass[aps,prl,amsmath,amssymb,twocolumn,preprintnumbers,nofootinbib,longbibliography,superscriptaddress,10pt]{revtex4-1}
\usepackage[utf8]{inputenc} 
\usepackage{graphicx}
\usepackage{url}
\usepackage[bookmarks, pagebackref=false]{hyperref}
\usepackage[usenames,dvipsnames]{xcolor}
\definecolor{orange}{cmyk}{0,0.5,1,0}
\definecolor{rossoCP3}{cmyk}{0,.88,.77,.40}
\definecolor{graa}{rgb}{0.8,0.8,0.8}
\definecolor{blaa}{rgb}{0.2,0.2,0.6}
\hypersetup{
  colorlinks,
  bookmarksopen,
  bookmarksnumbered,
  citecolor=blaa, 		
  linkcolor=rossoCP3,	
  urlcolor=rossoCP3,			
}
\usepackage{bm}
\usepackage{pxfonts}
\usepackage{amsmath}
\usepackage{amssymb}
\usepackage{amsfonts}
\usepackage{graphicx}
\usepackage{amssymb}
\usepackage{dsfont}%
\usepackage{slashed}
\usepackage{braket}
\usepackage{natbib}

\usepackage{adjustbox}
\usepackage{tikz}
\usetikzlibrary{patterns}
\usetikzlibrary{decorations.pathmorphing}
\usetikzlibrary{decorations.markings}
\usepgflibrary{arrows.meta}
\tikzstyle{box}=[pattern=north west lines, pattern color=colD, draw opacity=0.5,fill opacity=0.2,draw=colD,rounded corners=10]
\tikzstyle{op}=[circle, draw=black, fill=white,inner sep=0pt,minimum size=0pt]
\tikzstyle{prop}=[-, draw=colA, line width=1pt]
\tikzstyle{legOp}=[-, draw=colA, line width=1pt]
\tikzstyle{legExt}=[-, draw=colA, line width=1pt]

\definecolor{colA}{HTML}{c19277}
\definecolor{colB}{HTML}{e1bc91}
\definecolor{colD}{HTML}{62959c}

\newcommand{\ag}{\ensuremath{a_e}}
\newcommand{\alam}{\ensuremath{\lambda}}

\newcommand{\be}{\begin{equation}}
\newcommand{\ee}{\end{equation}}
\newcommand{\beq}{\begin{eqnarray}}
\newcommand{\eeq}{\end{eqnarray}}

\newcommand{\p}{{\cal P}\exp}

\newcommand{\bmp}{\noindent\begin{minipage}{16cm}}
\newcommand{\emp}{\end{minipage}\vskip 7mm} 

\renewcommand \l  {\lambda}

\newcommand*{\del}{\mathop{\mathrm{{}\partial}}\mathopen{}}

\newcommand{\MSbar}{$\overline{\textrm{MS}}$}
\renewcommand{\d}{\textrm{d}}

\def\lsim{\mathrel{\rlap{\lower4pt\hbox{\hskip1pt$\sim$}}
    \raise1pt\hbox{$<$}}}                
\def\gsim{\mathrel{\rlap{\lower4pt\hbox{\hskip1pt$\sim$}}
    \raise1pt\hbox{$>$}}}                

\def\r{\rho}
\def\a{\alpha}

\def\G{\Gamma}
\def\d{\delta}
\def\D{\Delta}
\def\e{\epsilon}

\def\p{\pi}

\def\m{\mu}
\def\n{\nu}

\def\l{\lambda}

\def\s{\sigma}

\def\IR{\relax{\rm I\kern-.18em R}}

\def\IR{\relax{\rm I\kern-.18em R}}
\def\IL{\relax{\rm I\kern-.18em L}}

\def\inv{^{\raise.15ex\hbox{${\scriptscriptstyle -}$}\kern-.05em 1}}

\begin{document}


\title{ Gauge Invariance at Large Charge}

\author{Oleg {\sc Antipin}}\email{oantipin@irb.hr}
\affiliation{Rudjer Boskovic Institute,
  Division of Theoretical Physics,
  Bijeni\v cka 54, 10000 Zagreb, Croatia}
\author{Alexander {\sc Bednyakov}}\email{bednya@theor.jinr.ru}
\affiliation{Bogoliubov Laboratory of Theoretical Physics,
  Joint Institute for Nuclear Research,
  Dubna 141980, Russia}
\author{Jahmall {\sc Bersini}}\email{jbersini@irb.hr}
\affiliation{Rudjer Boskovic Institute,
  Division of Theoretical Physics,
  Bijeni\v cka 54, 10000 Zagreb, Croatia}
  \affiliation{Kavli IPMU (WPI), UTIAS, The University of Tokyo, Kashiwa, Chiba 277-8583, Japan}
\author{Pantelis  {\sc Panopoulos}}\email{Pantelis.Panopoulos@irb.hr}
\affiliation{Rudjer Boskovic Institute,
  Division of Theoretical Physics,
  Bijeni\v cka 54, 10000 Zagreb, Croatia}
\author{Andrey {\sc Pikelner}}\email{pikelner@theor.jinr.ru}
\affiliation{Bogoliubov Laboratory of Theoretical Physics,
  Joint Institute for Nuclear Research,
  Dubna 141980, Russia}
\affiliation{The Institute of Nuclear Physics,
  Ministry of Energy of the Republic of Kazakhstan,
  Almaty, 050032, Kazakhstan}

\begin{abstract}
  Quantum field theories with global symmetries simplify considerably in the
  large-charge limit allowing to compute correlators via a semiclassical expansion
  in the inverse powers of the conserved charges. A generalization of the approach
  to gauge symmetries has faced the problem of defining gauge-independent
  observables and, therefore, has not been developed so far. We employ the
  large-charge expansion to calculate the scaling dimension of the lowest-lying
  operators carrying $U(1)$ charge $Q$ in the critical Abelian Higgs model in $D=
  4-\e$ dimensions to leading and next-to-leading orders in the charge and
  all orders in the $\e$ expansion. Remarkably, the results match our independent
  diagrammatic computation of the three-loop scaling dimension of the operator
  $\phi^Q(x)$ in the Landau gauge. We argue that this matching is a consequence of
  the equivalence between the \emph{gauge-independent} dressed two-point function of
  Dirac type with the \emph{gauge-dependent} two-point function of $\phi^Q(x)$ in the
  Landau gauge. We, therefore, shed new light on the problem of defining
  gauge-independent exponents which has been controversial in the literature on
  critical superconductors as well as lay the foundation for large-charge methods
  in gauge theories.
\end{abstract}
\maketitle
\small
\section{Introduction and Overview}

The field theoretical description of condensed matter systems is one of the
pillars of contemporary physics. In this context, superconductors are defined as
the materials with $U(1)$ gauge invariance in the broken phase with the relevant
field theory taking the name of Abelian Higgs model. This is a very well-known
textbook example of the Higgs mechanism and is relevant to the description of a
plethora of physical systems such as the aforementioned superconductors
\cite{Ginzburg:1950sr, herbut_2007, Folk, Dasgupta:1981zz}, liquid crystals
\cite{Halperin:1973jh}, cosmic strings \cite{Hindmarsh:2008dw}, and vortex lines
in superfluids \cite{Kleinert:1989kx}. However, one of the main issues regarding this model is the definition
of gauge invariant correlation functions describing physical quantities. Consider as a starting point the two-point function of a
complex scalar field $G_{\phi}(x_f-x_i)\equiv
\langle\bar\phi(x_f)\phi(x_i)\rangle$; while this correlator is invariant under
global $U(1)$ transformations, it violates $U(1)$ gauge symmetry and thus it
vanishes identically due to Elitzur's theorem \cite{Elitzur:1975im}. To make
progress, one is led to define a gauge invariant generalization of this
correlator which, however, is not unique and different approaches were shown to
lead to different physical results. In particular, the two main proposals
considered in the literature date back to the works of Dirac
\cite{dirac1981principles} and Schwinger \cite{Schwinger:1962tp,
Schwinger:1959xd}. Dirac's approach introduces a Wilson line as
\be
 G_D(x_f-x_i)=\langle \bar\phi(x_f)\exp\left(i \ e\int d^D x J^\mu (x) A_\mu(x)\right)\phi(x_i)\rangle
 \label{Dirac}
\ee
where $\del^{\m}J_{\m}=\d(x-x_f)-\d(x-x_i)$ and $\del^2J_{\m}=0$. The explicit form of the non-local current is $J_\mu = J^{'}_{\m}(z- x_f) - J^{'}_{\m}(z- x_i)$ with
\be
J^{'}_{\m}(z) =-i \int \frac{d^D k}{(2 \pi)^D} \frac{k_\mu}{k^2} e^{i k \cdot z}=- \frac{\Gamma(D/2-1)}{4 \pi^{D/2}}\del_\mu \frac{1}{z^{D-2}}\,\,.
\ee
Based on the definitions above $G_D(x_f-x_i)=\langle  \bar \phi_{nl}(x_f) \phi_{nl}(x_i)\rangle$ where
\be\label{order}
\phi_{nl}(x)\equiv e^{-ie\int d^Dz J^{'}_{\m}(z-x)A^{\mu}(z)}\phi(x)
\ee
has been proposed as the {\it non-local order parameter} for the superconducting
phase transition. Noticeably, in the Landau gauge $\del^{\m}A_{\m}=0$,
$\phi_{nl}(x)$ reduces to the {\it local} order parameter $\phi(x)$ since $J^{'}_{\m}$
is a total derivative \cite{Kennedy:1985yn}. Therefore, we expect that the
result for \emph{gauge-independent} Dirac correlator will coincide with the
result obtained for the \emph{gauge-dependent} $G_{\phi}(x_f-x_i)$ correlator in
the Landau gauge. Physically, $\phi_{nl}(x)$ can be interpreted as the creation
operator of a charged scalar particle dressed with a coherent state of photons
describing its Coulomb field. On the other hand, in the Schwinger approach the
gauge invariant correlator reads
\be
G_S(x_f-x_i)=\langle \bar \phi(x_f)\exp\left(-ie\int d x^{\m}
  A_\mu(x)\right)\phi(x_i)\rangle\,\,
\ee
and, analogously to its Dirac counterpart, $G_S(x_f-x_i)$ reduces to
$G_\phi(x_f-x_i)$ in the \emph{traceless gauge} \cite{Kleinert:2003}.

Of special interest are the correlators computed in the critical model, where
they take the form:
\be
G_{\phi,D,S}(x_f-x_i)=\frac{1}{\rvert x_f-x_i \rvert^{D-2+2\eta_{\phi,D,S}}} \ .
\ee
In the context of the $O(n)$ model, it has been verified in
\cite{Kleinert:2005sa} (\cite{Kleinert:2003}) that $\eta_D$ ($\eta_S$) coincides
with $\eta_{\phi}$ in the Landau (traceless) gauge to leading order in the
$\epsilon$ expansion and to leading order in $1/n$. Moreover, it has been
pointed out in \cite{Frohlich:1981yi}, that $G_{S}(x_f-x_i)$ cannot provide a
correct description of the model in the broken phase since it does not lead to
long-range order.

In this Letter, we study the issue of extracting gauge-invariant critical
exponents from the novel perspective of the large-charge expansion which has
attracted a lot of attention recently \cite{Hellerman:2015nra, Gaume:2020bmp,
Badel:2019oxl, Antipin:2020abu, Monin:2016jmo, Alvarez-Gaume:2016vff,
Banerjee:2017fcx, Giombi:2022gjj, Antipin:2022naw}. There, one starts with a
critical theory invariant under a set of global symmetries and evaluates scaling
exponents via an expansion in inverse powers of the conserved charges. In
practice, this is achieved by computing the ground state energy at fixed charge
$Q$ of the theory defined on the unit cylinder $\mathbb{R}\times S^{D-1}$,
which, by virtue of the state-operator correspondence
\cite{Cardy:1984rp,Cardy:1985lth} is equal to the scaling dimension $\Delta_Q$
of the lowest-lying operators with charge $Q$. In perturbative models, the
approach amounts to a semiclassical evaluation of the expectation value of the
evolution operator $e^{-HT }$ in an arbitrary state with fixed charge $Q$ which
allows to resum an infinite number of Feynman diagrams at every order of the
semiclassical expansion \cite{Badel:2019oxl, Antipin:2020abu}. At first sight, a
generalization of the approach to gauge theories looks impossible due to the
lack of gauge invariance of correlators of matter fields such as $G_{\phi}(x_f-x_i)$. Remarkably, we instead
show by explicit calculation that our results for $\Delta_Q$ in the large-charge
expansion are gauge-independent and correspond to the conformal dimensions of
$\phi^Q(x)$ computed in Landau gauge. We interpret the result as the consequence of the equivalence in
the Landau gauge of $\phi^Q(x)$ to the non-local operator ${\phi}_{nl}^Q(x)$
generalizing the Dirac construction \eqref{order} to arbitrary values of $Q$.

Ultimately, our results confirm that the critical exponent $\eta_S$ is not
adequate to describe the broken phase. In fact, $\eta_S$ is more negative than
$\eta_D$ at the leading order in the $\e$ expansion \cite{Kleinert:2005sa,
Kleinert:2003}. Therefore, it leads to a lower conformal dimension which, in
turn, would be computed in our approach by construction.

Highlighting our findings:
\begin{itemize}
\item We show that the large-charge expansion can be applied also to gauge
  theories where the relevant gauge-invariant observables are in general
  non-local.

\item We explicitly show that the non-local operators ${\phi}_{nl}^Q$ are the
  lowest-lying operators with charge $Q$ well-defined at criticality. In
  particular, this signals that ${\phi}_{nl}$ is the relevant order parameter for
  long-range order in superconductors and it is automatically selected by the large charge
  approach.

\item We compute $\Delta_Q$ to the next-to-leading order in the large-charge
  expansion and all orders in the loop expansion. Moreover, we explicitly
  calculate the full three-loop $\Delta_Q$ in perturbation theory and find perfect
  agreement with our semiclassical result.
\end{itemize}

\section{Large charge operators in Scalar QED}
\label{gaugemodel}

\subsection{Classical contribution}

Our starting point is the massless gauged $(\bar\phi\phi)^2$ theory in $D=4-\e$ dimensions given by the action
\be
\label{gauge}
S=\int d^D x\left(\frac{1}{4}F_{\m\n}F^{\m\n}+\left(D_{\m}\phi\right)^{\dag}D_{\m}\phi+\frac{\l (4 \pi)^2}{6}(\bar\phi\phi)^2\right) \,,
\ee
where $D_{\m}\phi=(\del_{\m}+ie A_{\m})\phi$ and $F_{\mu \nu} = \del_\mu A_\nu -
\del_\nu A_\mu$. The above action is invariant under the local $U(1)$ gauge
transformations $\phi\to e^{i\a(x)}\phi$ and $A_{\m}\to
A_{\m}-e^{-1}\del_{\m}\a(x)$. The equations of motion (EOM) are given by
\be
\begin{split}\label{eom}
-D^{\m}D_{\m}\phi+\frac{\l  (4 \pi)^2}{3}\,(\bar\phi\phi)\bar\phi=0\,, \qquad
 \del_{\m}F^{\m\n}=J^{\n} \,,
\end{split}
\ee
where $J_{\m}$ is the $U(1)$ electromagnetic current. The associated charge is
defined as $Q = \int d^{D-1}x J^0$. Note that since the $U(1)$ symmetry is
gauged, in order to avoid long-range electric fields causing infrared
divergences the system needs to be electrically neutral. This can be achieved by
introducing a neutralizing background $J^{\nu}_{b}$ such that $J_{total}^{0}=
J^{0}-J^{0}_{b}=0$. This background current may be seen as the one
used to define the non-local operators ${\phi}_{nl}^Q$. However, being non-dynamical, it does not affect the scaling dimension $\D_Q$. Throughout the text
we use the same notation for bare and renormalized couplings. The theory
features a Wilson-Fisher fixed point which, at the one-loop level, reads
\cite{Coleman:1973jx, herbut_2007}
\begin{align}
    \l^* = \frac{3}{20} \left(19 \epsilon \pm i \sqrt{719} \epsilon \right) \,, \qquad \ag^{*} = \frac{3}{2} \epsilon \ ,
    \label{FP}
\end{align}
with $\ag=\frac{e^2}{(4\pi)^2}$. Note that the fixed point occurs at complex
values of $\l$. However, once $\Delta_Q$ is calculated at the fixed point, one
can rewrite the result in terms of the renormalized couplings and obtain an
expression valid for arbitrary values of $\l$ and $\ag$ \cite{Antipin:2020rdw}.
Mapping the theory to the cylinder $\mathbb{R}\times S^{D-1}$ with unit radius,
the Weyl invariant action becomes
\be
\label{phiA}
S=\int d^D x\,\sqrt{-g}\left(\frac{1}{4} F_{\m\n} F^{\m\n}+\left( D_{\m}\phi\right)^{\dagger}D^{\m}\phi+m^2\,\bar\phi\phi+\frac{\l   (4 \pi)^2}{6}(\bar\phi\phi)^2\right) \,,
\ee
The scalar mass arises from the conformal coupling between $\phi$ and the Ricci
scalar of the $S^{D-1}$ sphere and is given by $m^2={(D-2)^2}/{4}$
\cite{Brown:1980qq}. Due to the state-operator correspondence, after
parametrizing the scalar field as
$\phi(x)=\frac{\rho(x)}{\sqrt{2}}e^{i\chi(x)}$, the calculation of the scaling
dimensions reduces to the evaluation of the following matrix element
\be\label{2point}
 \langle Q| e^{-HT }| Q\rangle=\mathcal Z^{-1}\int_{\rho=f}^{\rho=f} \mathcal D\rho\mathcal D\chi  \mathcal D A \,e^{-S_{\text{eff}}} \,,
\ee
where the effective action $S_{\text{eff}}$ is succinctly written as
\be
\label{effrhochiA}
\begin{split}
&S_{\text{eff}}=\int_{-T/2}^{T/2}d\tau\int d\Omega_{D-1}   \Bigg(\,\frac{1}{4} F_{\m\n} F^{\m\n}+\frac{1}{2}(\del\rho)^2+\frac{1}{2}\rho^2(\del\chi)^2\\
&+\frac{1}{2}{m^2}\rho^2
+e\rho^2A_{\m}\del^{\m}\chi+\frac{1}{2}e^2\rho^2A_{\m}A^{\m}+\frac{\l  (4 \pi)^2}{24}\rho^4+\frac{i\,Q}{\Omega_{D-1}}\dot\chi\Bigg) \,.
\end{split}
\ee
Here $\mathcal Z=\int \mathcal D \phi\mathcal D\bar\phi \mathcal D A\,e^{-S} $
is the partition function, $\ket{Q}$ is an arbitrary state with charge $Q$, and
$f$ is a fixed value for the $\rho(x)$ field. The last term fixes the charge of
initial and final states. More details on the approach can be found in e.g.
\cite{Badel:2019oxl, Antipin:2022naw}. The above matrix element is related to $\Delta_Q$ as 
\be\label{correl}
 \langle Q| e^{-HT }| Q\rangle \underset{T \to \infty}{=} \mathcal{N} e^{- \Delta_Q T}  \,,
\ee
where $\mathcal{N}$ is an unimportant T-independent normalization coefficient. By rescaling the fields as $\rho \to \sqrt{Q} \rho$, $A_\mu \to \sqrt{Q} A_\mu$ to exhibit $Q$ as the loop counting parameter, one sees that  \eqref{2point} can be calculated via a
semiclassical expansion around the saddle points of $S_{\text{eff}}$. Accordingly, $\Delta_Q$ takes the following form
\be
\Delta_Q = \sum_{j=-1}^{\infty} \frac{1}{Q^j}\Delta_j \left(Q e^2, Q \lambda\right) \ .
\label{expansion}
\ee
Every $\Delta_j$ resums an infinite series of Feynman diagrams of the
conventional perturbative expansion which can be recovered by expanding the
$\D_j$ for small values of the 't Hooft-like couplings $Q e^2$ and $Q \lambda$.
The lowest energy solution of the EOM corresponds to a homogeneous ground state
\be\label{rhochi}
\rho(x)=f,\quad \chi(x)=-i\m\tau,\quad A_{\m}=0\,\,.
\ee
The vev of the radial mode $f$ and the chemical potential $\mu$ are determined
in terms of $Q$ and the couplings via the EOM as
\be
\label{chem}
\m^3-\m=\frac{4}{3}\l Q\,\,, \qquad f^2 = \frac{6}{(4 \pi)^2 \lambda} \left(\mu^2 - m^2\right)\,\,.
\ee
The classical contribution to $\Delta_Q$ is then obtained by plugging the
solution \eqref{rhochi} into $S_{\text{eff}}$. Since at the classical level
$A_\mu =0$, the leading order of the semiclassical expansion is equivalent to
the one obtained in \cite{Badel:2019oxl} for the ungauged $(\bar\phi \phi)^2$
model, reading
 \be
 \label{classic}
 4\,\Delta_{-1}=  \frac{3^{2/3}\left(x+\sqrt{-3+x^2}\right)^{{1}/{3}}}{3^{1/3}+\left(x+\sqrt{-3+x^2}\right)^{{2}/{3}}}  + \frac{3^{1/3}\left(3^{1/3}+\left(x+\sqrt{-3+x^2}\right)^{{2/3}}\right)}{\left(x+\sqrt{-3+x^2}\right)^{{1/3}}}
\ee
where $x \equiv 6 \lambda Q$.

\subsection{Leading quantum correction}
The next-to-leading order $\D_0$ is given by the functional determinant of the
fluctuations around the classical solution and can be expressed as a sum of
zero-point energies
\begin{equation}
  \label{eq:one-loop-det1}
  \Delta_0 = \frac{1}{2}\sum_{\ell=\ell_0}^\infty \sum_i d_\ell \ \omega_i(\ell)\,\,,
\end{equation}
where the innermost sum runs over all the dispersion relations of the spectrum, $d_\ell$ is the degeneracy of the corresponding eigenvalues of the momentum and $\ell_0$ is given in Table \ref{table1}.
To find the spectrum, we expand the fields around the classical configuration
\eqref{rhochi} as $\r(x)=f+r(x)$ and $\chi(x)=-i\m\tau+ f^{-1}{\pi(x)}$. The
action at the quadratic order in the fluctuations takes the form
\be
\label{Higgs}
\begin{split}
S_{\text{eff}}^{(2)}=\int_{-T/2}^{T/2}d\tau&\int d\Omega_{D-1}\,\Big(\,\,\frac{1}{4}F_{\m\n}^2+\frac{1}{2}(\del_{\m} r)^2+\frac{1}{2}(\del_{\m}\pi)^2\\
&-\frac{1}{2}\,2(m^2-\m^2)r^2-2i\m r\del_{\tau}\pi+ef\del_{\m}\pi A^{\m}\\
&-2ie\m f rA_0+\frac{1}{2}(ef)^2A_{\m}A^{\m}\Big)\,.
\end{split}
\ee
It is clear that the Higgs mechanism has occurred with the gauge field acquiring
mass $m_A^2=(ef)^2$ and $\p(x)$ being the massless Goldstone mode of the
spontaneously broken $U(1)$ symmetry. Nevertheless, due to Elitzur's theorem
\cite{Elitzur:1975im} the local part of a gauge symmetry of a compact group
cannot be spontaneously broken and the action \eqref{Higgs} is invariant under
the residual gauge symmetry
\be
\label{res}
\d r=0\,,\qquad \d\pi=f\a(x)\,,\qquad \d A_{\m}=-\frac{1}{e}\,\del_{\m}\a(x) \,,
\ee
where $\a(x)$ is the phase of the original $U(1)$ gauge transformations. To
evaluate the path integral we employ the $R_{\xi}$-gauge and the Faddeev-Popov
method. Then the action \eqref{Higgs} is replaced as $S_{\text{eff}}^{(2)} \to
S_{\text{eff}}^{(2)} +\frac{1}{2}\int d^D x G^2 \,,$ with
\be \label{23}
G=\frac{1}{\sqrt\xi}
\left(\nabla_{\m}A^{\m}+ e f \p\right)\,\,, \qquad \frac{\d G}{\d \a}=\frac{1}{\sqrt\xi}\left(-\frac{1}{e}
\nabla^2+ ef^2\right)\,\,,
\ee
where we used \eqref{res} to take the variation with respect to $\alpha$. Rescaling
$G$ with $e$, the determinant of $\d G/\d\a$ can be represented using a set of
Fadeev-Popov ghosts $\bar c,c$ with
$
\label{ghost}
\mathcal L_{\text{ghost}}=\bar c\left(-\nabla^2+(ef)^2\right)c\,.
$
Consequently, the quadratic Lagrangian becomes
\be
\label{Lquad}
\begin{split}
\mathcal L_{\text{eff}}^{(2)}&=\frac{1}{2}A_{\m}\left(-g^{\m\n}\nabla^2+\mathcal R^{\m\n}+\left(1-\frac{1}{\xi}\right)\nabla^{\m}\nabla^{\n}+(e f)^2g^{\m\n}\right)A_{\n}\\
&+\frac{1}{2}(\del_{\m}r)^2-\frac{1}{2}\,2(m^2-\m^2)r^2+\frac{1}{2}(\del_{\m}\p)^2-\frac{1}{2\xi}(ef)^2\p^2\\
&-2i\m \,r\del_{\tau}\p-2if\m rA^0+ef\left(1-\frac{1}{\xi}\right)A_{\m}\del^{\m}\p+\mathcal L_{\text{ghost}} \,\,,
\end{split}
\ee
where we used that $[\nabla^{\m},\nabla_{\n}]A_{\mu}=\mathcal R^{\m}_{\n}A_{\m}$
with $\mathcal R^{\m}_{\n}=(D-2)\,\d^{\m}_{\n}$ the Ricci tensor on $S^{D-1}$.
With these results at our disposal, we are ready to evaluate the determinant of
the partition function using
$-\nabla^2=-\del_{\tau}^2+\left(-\nabla^2_{S^{D-1}}\right)$ on $\mathbb{R}\times
S^{D-1}$ space. The sphere Laplacian eigenvalues and degeneracies are shown in
Table.\ref{table1}.

\begin{table*}
  \begin{center}
    \begin{tabular}{ccccc}\hline\hline
      Field & $d_\ell$ & $\omega_i (\ell)$ &  $\ell_0$ \\
      \hline
      $B_i$ & $n_v(\ell)$ &  $\sqrt{J_{\ell(v)}^2 +(D-2) + e^2 f^2}$ &1 \\
      $C_i$ & $n_s(\ell)$ &  $\sqrt{J_{\ell(s)}^2+ e^2 f^2}$ &1 \\
      $(c,\bar c)$ & $-2 n_s(\ell)$ & $\sqrt{J_{\ell(s)}^2+ e^2 f^2}$ &0 \\
      $A_0$ &  $n_s(\ell)$ &  $\sqrt{J_{\ell(s)}^2+ e^2 f^2}$ & 0\\
      $\phi$ & $n_s (\ell)$ & $\sqrt{J_{\ell(s)}^2+3 \mu ^2-m^2 +\frac{1}{2} e^2 f^2  \pm \sqrt{\left(3 \mu ^2-m^2-\frac{1}{2} e^2 f^2\right)^2+4 J_{\ell(s)}^2 \mu ^2}}$ & 0\\
      \hline\hline
    \end{tabular}
  \end{center}
  \caption{\small The fluctuations spectrum of the model. The second column shows the scalar and vector fields degeneracies, where $\displaystyle{n_s(\ell)={(2\ell+D-2)\G(\ell+D-2)}/{\G(D-1)\G(\ell+1)}}$ and  $n_{v}(\ell)={\ell(\ell+D-2)(2\ell+D-2)\G(\ell+D-3)}/{\G(\ell+2)\G(D-2)}$. The third column contains the dispersions entering \eqref{eq:one-loop-det1}, where $J_{\ell(s)}^2=\ell(\ell +D-2)$ and
	  $J_{\ell(v)}^2=\ell\left(\ell+D-2\right)-1$ are the Laplacian eigenvalues on $S^{D-1}$.       }
  \label{table1}
\end{table*}
Let us separately comment on the first line of \eqref{Lquad}. When the operator
in the brackets acts on $A_0$, which is a scalar on $S^{D-1}$, the Ricci tensor
does not contribute and $-\nabla^2$ is equivalent to the scalar Laplacian. On
the other hand, the vector field $A^i$ is decomposed as the kernel plus the
image of the nabla-operator. In mathematical terms $A^i=B^i+C^i$, where
$\nabla_iB^i=0$ and $C^i=\nabla^i h $ with $h$ an arbitrary function. These
fields are orthogonal to each other and terms containing products of them
vanish. This implies that $B^i$ is a vector while the $C^i$ are a set of
scalars. As a consequence, the $A_0$ and $C_i$ fields can be organized in the
same scalar multiplet while the Gaussian path integration over the $B^i$
evaluates to

\be\label{detB}
\begin{split}
\int \frac{d\omega}{2\p}\sum_{\ell}n_v(\ell)\det\left(-\del_{\tau}^2+J_{\ell(v)}^2+(D-2)+(ef)^2\right)^{-1/2} \,.
\end{split}
\ee
In addition, using \eqref{Lquad} along with the scalar components of $A^\mu$, we
obtain the inverse propagator matrix for the $(r,\pi,A_0,C_i)$ fields
\begin{widetext}
\be\label{matrix}
{\cal B} =\left(
  \begin{array}{cccc}
   - \omega^2+J_{\ell(s)}^2+2(\m^2-m^2) & -2i\m\omega   &-2ie\m f&0 \\
  2i\m \omega & -\omega^2+J_{\ell(s)}^2+\frac{1}{\xi}e^2f^2 & -ef\left(1-\frac{1}{\xi}\right)\omega & -i ef\left(1-\frac{1}{\xi}\right)|J_{\ell(s)}|\\
   -2ie\m f  & ef\left(1-\frac{1}{\xi}\right)\omega & -\frac{1}{\xi}\omega^2+J_{\ell(s)}^2+(ef)^2 & i\left(1-\frac{1}{\xi}\right) \omega|J_{\ell(s)}|\\
   0     & ief\left(1-\frac{1}{\xi}\right) |J_{\ell(s)}|&
 i\left(1-\frac{1}{\xi}\right)\omega|J_{\ell(s)}| &-\omega^2+\frac{1}{\xi}J_{\ell(s)}^2+(ef)^2 \\
    \end{array}
\right) \,\,.
\ee
\end{widetext}
The determinant of \eqref{matrix} factorizes as
\be
\label{xidet}
\xi\det\mathcal B=(\omega^2+\omega_+^2)(\omega^2+\omega_-^2)(\omega^2+\omega_1^2)^2 \ .
\ee
Table \ref{table1} includes the dispersion relations $\omega_{\pm}, \omega_1$
along with the ones from the $B^i$ fields in \eqref{detB} and the ghosts. The
functional determinant of the ghosts cancels against the contribution stemming
from $C_i$ and $A_0$, leaving a single ghost zero mode ($\ell=0$) contribution
which, in turn, cancels one of the scalar zero modes. Also, since in
\eqref{2point} the quadratic part of the partition function $\mathcal Z$ in the denominator contains the  gauge field, we should gauge fix it as well. Following the same gauge fixing procedure as above and performing the Gaussian integration, the $\xi$-dependence factorizes as in \eqref{xidet} and cancels out in the final result. The obtained expression \eqref{eq:one-loop-det1} arising from the Gaussian integrals in the numerator and the denominator is divergent and needs to be renormalized. To this end, we follow \cite{Badel:2019oxl} and regularize the sum over $\ell$ by subtracting the divergent terms in the $\ell \to \infty$ limit. Then we use dimensional regularization to isolate the $\frac{1}{\epsilon}$ pole which is canceled by performing renormalization in the usual \MSbar~scheme. The renormalized result reads
\begin{widetext}
  \begin{align}\label{D0}
    \Delta_0 & =\,\, \frac{1}{16} \left(-15 \mu ^4-6 \mu ^2+8 \sqrt{6 \mu ^2-2}+5\right) + \frac{1}{2} \sum_{\ell=1} \sigma (\ell)
               -\frac{3\ag}{8\alam}(\mu^2-1)\left(\frac{3\ag}{\alam}(7\mu^2+5) -9 \mu^2 + 5\right) \,,
  \end{align}
  where the summand $\s(\ell)$ is given by
  \begin{align}
    \sigma(\ell) &=
                 \frac{9\ag}{2\alam \ell} \left(\mu^2-1\right)\left[\left(\frac{3\ag}{\alam}-1\right)\left(\mu^2-1\right)-2 \ell(\ell+1)\right]
                 +\frac{5}{4 \ell} \left(\mu^2-1\right)^2
                 - 2 (\ell + 1) ( 2 \ell (\ell + 2) + \mu^2)
                 + (\ell+1)^2 \left(R_{+} + R_{-}\right) +2 \ell (\ell+2) R
                 \,\,,
\end{align}
with
 \begin{equation}
 \label{eq:roots-def}
     R_{\pm} = \sqrt{\frac{3 \ag}{\alam}\left(\mu ^2-1\right)+3 \mu ^2+\ell (\ell+2)-1
       \pm\sqrt{\left(\frac{3 \ag }{\alam}\left(\mu ^2-1\right)-3 \mu ^2+1\right)^2+4 \ell (\ell+2) \mu ^2}},
     \quad
     R=\sqrt{\frac{6 \ag  }{\alam} \left(\mu ^2-1\right) +\ell (\ell+2)+1}\,\,.
 \end{equation}
\end{widetext}
Eq.\eqref{D0} is our main result which, combined with \eqref{classic}, gives
$\Delta_Q$ to the next-to-leading order in the large-charge expansion and all
orders in the loop expansion at the fixed point \eqref{FP}.

\section{Explicit three-loop calculation}
\label{sec:calc-3l}
In perturbation theory we are able to compute the anomalous dimension of
$\phi^Q$ operator as a series in small couplings $\ag$ and $\alam$
\begin{equation}
  \label{eq:Delta-pert}
  \gamma_Q(\alam,\ag,\xi) = \sum\limits_{l=1}^\infty \gamma_Q^{(l-\text{loop})}(\alam,\ag,\xi)\ , \ \  \gamma_Q^{(l-\text{loop})} \equiv \sum_{k=0}^{l} C_{kl} Q^{l+1-k}
\end{equation}
where we use the linear $R_\xi$ gauge. The $l$-loop contribution
$\gamma_Q^{(l-\text{loop})}$ is a polynomial of degree $l+1$ in $Q$ and to find
the coefficients $C_{kl}$ at $l$ loops one can explicitly compute the anomalous
dimensions of $\phi^Q$ operators for fixed $Q=1,\ldots,l+1$.

The case $Q=1$ corresponds to the well-known field anomalous dimension. 
In addition, we considered one-particle irreducible Green functions with $\phi^Q$ operator insertions (see Fig.~\ref{fig:ints3l}) for $Q=2,3,4$. We used the infrared rearrangement trick
\cite{Vladimirov:1979zm} together with \texttt{MATAD} \cite{Steinhauser:2000ry} package
to find the required renormalization constants $Z_Q$ in the \MSbar~scheme at
three loops. This allows us to derive the following expression for the anomalous
dimension at arbitrary $Q$:
\begin{figure*}[t]
  \centering
  \begin{tabular}[t]{cccccc}
    \adjustbox{valign=c}{\resizebox{2.5cm}{!}{
		    \includegraphics{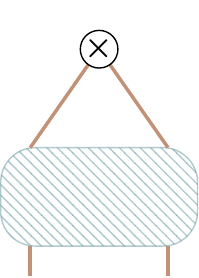}
    }}

    &
      \adjustbox{valign=c}{\resizebox{2.5cm}{!}{
		    \includegraphics{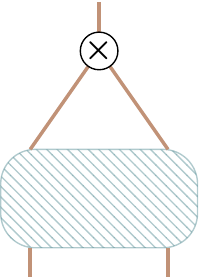}
      }}

    &
      \adjustbox{valign=c}{\resizebox{2.5cm}{!}{
		      \includegraphics{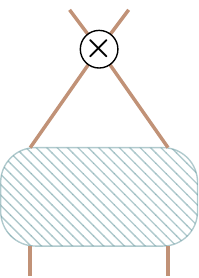}
      }}

    &
      \adjustbox{valign=c}{\resizebox{2.5cm}{!}{
		      \includegraphics{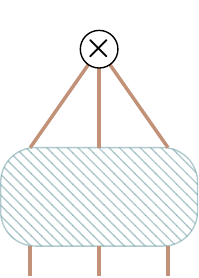}
      }}

    &
      \adjustbox{valign=c}{\resizebox{2.5cm}{!}{
		      \includegraphics{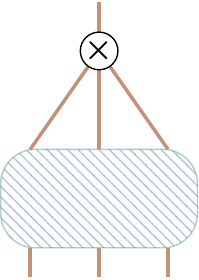}
      }}

    &
      \adjustbox{valign=c}{\resizebox{2.5cm}{!}{
		      \includegraphics{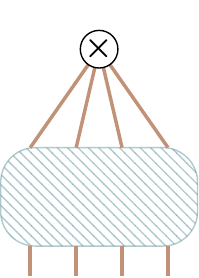}
      }}

    \\
    \texttt{2-2} & \texttt{3-2} & \texttt{4-2} & \texttt{3-3} & \texttt{4-3} & \texttt{4-4}
  \end{tabular}
  \caption{Operator insertions into various Green's functions considered up to the
    three-loop order. Abbreviation $Q-M$ means insertion of the operator
    $\phi^Q$ with $M$ legs attached to the loop diagram. $M=2$ contribution starts
    from the 1-loop order, $M=3$ from two-loop and $M=4$ from three-loop order
    respectively. Hatched blobs include loop corrections calculated with standard
    scalar QED Feynman rules.}
  \label{fig:ints3l}
\end{figure*}
{
  \footnotesize
  \begin{align}
    &    \gamma_Q^{(1)}(\alam,\ag,\xi)  =
      \underbrace{ \frac{\alam}{3} Q^2}_{\rm leading}
      - \underbrace{Q\left(3 \ag + \frac{\alam}{3}\right)}_{\rm sub-leading} + \ag Q^2 \xi,
      \label{eq:gamma_12}
      \\
     &\gamma_Q^{(2)}(\alam,\ag)  =
      \underbrace{- \frac{2 \alam^2}{9} Q^3}_{\rm leading}
      + \underbrace{\left(\ag^2 - \frac{4 \ag \alam}{3} + \frac{2 \alam^2}{9} \right) Q^2}_{\rm sub-leading}
      + \left(\frac{7 \ag^2}{3} + \frac{4 \ag \alam}{3} + \frac{\alam^2}{9}\right) Q,
      \nonumber
  \end{align}
}
\begin{widetext}
  \footnotesize
  \begin{align}
    &  \gamma_Q^{(3)}(\alam,\ag)  =
      \underbrace{\frac{8 \alam^3}{27} Q^4}_{\rm leading}
      + \underbrace{Q^3 \left[\frac{4 \ag \alam^2}{3} \left(3 - 2 \zeta_3\right)
      - \frac{8 \ag^2 \alam}{3}  \left(1 + 3 \zeta_3 \right)
      + 4 \ag^3 (9 \zeta_3 - 1) + \frac{2 \alam^3}{27} (16 \zeta_3 - 17)\right]}_{\rm sub-leading}
      \nonumber\\
    &   + Q^2 \left[ \frac{29 \ag^2 \alam}{6} + \ag^3 (95 - 108 \zeta_3) + \frac{\alam^3}{18} (57 - 64 \zeta_3) -
      \frac{4 \ag \alam^2}{9} (31 - 30 \zeta_3)\right]
      + Q \left[\frac{13 \ag^2 \alam}{6} + \frac{2 \ag \alam^2}{9} \left(49 - 48 \zeta_3\right) -
      \frac{2 \alam^3}{27} \left(31 - 32 \zeta_3\right) - \ag^3\left(\frac{3251}{54} - 72 \zeta_3\right)\right]
    \label{eq:gamma_3}
  \end{align}
\end{widetext}
As we argued above, we expect that once we fix the Landau gauge ($\xi=0$), which
corresponds to the fixed point of $\xi$ \cite{deCalan:1999vx}, and evaluate dimensions of operators
$\phi^Q$ at $\alam=\alam^*$, and $\ag=\ag^*$ \eqref{FP}, we should reproduce
\eqref{expansion}. We computed these scaling dimensions
\begin{align}
\Delta_Q =Q\left(\frac{D-2}{2}\right) + \gamma_Q(\alam^*,\ag^*,\xi^* = 0) \,,
\label{eq:DeltaQ_pert}
\end{align}
and indeed find perfect agreement with the corresponding terms from
\eqref{classic} and \eqref{D0} which, for reader's convenience, we highlighted
with brace below each term. In fact, as it was noticed in
Ref.~\cite{Jack:2021ypd, Antipin:2020abu}, at \emph{two loops and higher} the
expansion of \eqref{expansion} in small $\ag Q$ and $\alam Q$ precisely coincides
with perturbative result, even away from the fixed point which we confirmed
explicitly. In particular, the coefficients $C_{kl}$ appear in the small $\ag Q$
and $\alam Q$ expansion of $\Delta_{k-1}$. As for the one-loop contribution to $\Delta_Q$ \eqref{eq:DeltaQ_pert}, 
we find perfect agreement with Eq.~\eqref{expansion} only at the fixed point \eqref{FP}. Let us also note that the case $\ag=0$ corresponds to the
$O(2)$-symmetric model, for which the full six-loop result is available
\cite{Bednyakov:2022guj}. 

To facilitate further comparison with perturbative
calculations, we provide explicit expression for the coefficients $C_{0l}$ and $C_{1l}$ appearing in \eqref{eq:Delta-pert} from $l=4$ to $l=6$:
\allowdisplaybreaks
\begin{align} 
		\label{cicci}
		C_{04}=-\frac{14}{27}  \lambda ^4\,,\quad   C_{05}= \frac{256 \lambda ^5}{243}\,,\quad   C_{06}= -\frac{572 }{243}\lambda ^6 \,,
\end{align}
\begin{align}
C_{14} & = 2 \ag^4 (41 \zeta_{3}-125 \zeta_{5}+17)-\frac{4}{3} \ag^3 \alam (134 \zeta_{3}-140 \zeta_{5}+25) \nonumber\\
	& +\frac{2}{3} \ag^2 \alam^2 (80 \zeta_{3}-80 \zeta_{5}+33)-\frac{8}{27} \ag \alam^3 (7 \zeta_{3}-40 \zeta_{5}+37) \nonumber\\
	& -\frac{2}{81} \alam^4 (77 \zeta_{3}+80 \zeta_{5}-142),\\ \nonumber \\
C_{15} & = -4 \ag^5 (224 \zeta_{3}-50 \zeta_{5}-441 \zeta_{7}+95) \nonumber\\
       & +\frac{8}{3} \ag^4 \alam (413 \zeta_{3}-50 \zeta_{5}-525 \zeta_{7}+167) \nonumber \\
       & -\frac{8}{9} \ag^3 \alam^2 (164 \zeta_{3}-315 \zeta_{5}-140 \zeta_{7}+152) \nonumber\\
       & -\frac{4}{27} \ag^2 \alam^3 (368 \zeta_{3}+420 \zeta_{5}-560 \zeta_{7}+217) \nonumber\\
       & +\frac{4}{81} \ag \alam^4 (156 \zeta_{3}-130 \zeta_{5}-560 \zeta_{7}+559) \nonumber\\
       & +\frac{2}{243} \alam^5 (476 \zeta_{3}+480 \zeta_{5}+448 \zeta_{7}-1179),
\end{align}
\begin{align}
C_{16} & = 2 \ag^6 (5662 \zeta_{3}+462 \zeta_{5}-3514 \zeta_{7}-7434 \zeta_{9}+2487) \nonumber \\
       & -\frac{16}{3} \ag^5 \alam (2956 \zeta_{3}-319 \zeta_{5}-1680 \zeta_{7}-2583 \zeta_{9}+1270) \nonumber \\
       & +\frac{4}{9} \ag^4 \alam^2 (14932 \zeta_{3}-8350 \zeta_{5}-10185 \zeta_{7}-6300 \zeta_{9}+6286) \nonumber\\
       & -\frac{16}{27} \ag^3 \alam^3 (2413 \zeta_{3}-1335 \zeta_{5}-910 \zeta_{7}-840 \zeta_{9}+650) \nonumber\\
       & +\frac{8}{81} \ag^2 \alam^4 (2645 \zeta_{3}+515 \zeta_{5}-490 \zeta_{7}-2520 \zeta_{9}+1421) \nonumber\\
       & -\frac{4}{243} \ag \alam^5 (1886 \zeta_{3}-46 \zeta_{5}-2464 \zeta_{7}-4032 \zeta_{9}+4633) \nonumber\\
       & -\frac{2}{729} \alam^6 (3294 \zeta_{3}+3202 \zeta_{5}+3360 \zeta_{7}+2688 \zeta_{9}-10063) \ .
\end{align}
\section{Conclusions}

In the present work we took the first step towards the application of the
large-charge expansion in gauge theories. We demonstrated that our semiclassical \emph{gauge-independent} results for $\Delta_Q$ can be
interpreted  as scaling dimensions of non-local operators ${\phi}_{nl}^Q(x)$ that extend the Dirac construction to arbitrary $Q$.
As a byproduct of our analysis, we confirmed that the adequate critical exponent to describe the broken phase is $\eta_D$. Our work opens new
avenues for the large-charge methods in application to the most general
Gauge-Yukawa theories such as the Standard Model of particle physics and its
extensions.

\vskip 1em

\begin{acknowledgments}
The work of P. Panopoulos was supported by the Croatian Science Foundation
Project "New Geometries for Gravity and Spacetime" (IP-2018-01-7615). We wish to
thank Simeon Hellerman, Alexander Monin, and Francesco Sannino for fruitful
discussions at various stages of this work.
\end{acknowledgments}


\bibliography{SQEDlargecharge}

\begin{thebibliography}{34}%
\makeatletter
\providecommand \@ifxundefined [1]{%
 \@ifx{#1\undefined}
}%
\providecommand \@ifnum [1]{%
 \ifnum #1\expandafter \@firstoftwo
 \else \expandafter \@secondoftwo
 \fi
}%
\providecommand \@ifx [1]{%
 \ifx #1\expandafter \@firstoftwo
 \else \expandafter \@secondoftwo
 \fi
}%
\providecommand \natexlab [1]{#1}%
\providecommand \enquote  [1]{``#1''}%
\providecommand \bibnamefont  [1]{#1}%
\providecommand \bibfnamefont [1]{#1}%
\providecommand \citenamefont [1]{#1}%
\providecommand \href@noop [0]{\@secondoftwo}%
\providecommand \href [0]{\begingroup \@sanitize@url \@href}%
\providecommand \@href[1]{\@@startlink{#1}\@@href}%
\providecommand \@@href[1]{\endgroup#1\@@endlink}%
\providecommand \@sanitize@url [0]{\catcode `\\12\catcode `\$12\catcode
  `\&12\catcode `\#12\catcode `\^12\catcode `\_12\catcode `\%12\relax}%
\providecommand \@@startlink[1]{}%
\providecommand \@@endlink[0]{}%
\providecommand \url  [0]{\begingroup\@sanitize@url \@url }%
\providecommand \@url [1]{\endgroup\@href {#1}{\urlprefix }}%
\providecommand \urlprefix  [0]{URL }%
\providecommand \Eprint [0]{\href }%
\providecommand \doibase [0]{http://dx.doi.org/}%
\providecommand \selectlanguage [0]{\@gobble}%
\providecommand \bibinfo  [0]{\@secondoftwo}%
\providecommand \bibfield  [0]{\@secondoftwo}%
\providecommand \translation [1]{[#1]}%
\providecommand \BibitemOpen [0]{}%
\providecommand \bibitemStop [0]{}%
\providecommand \bibitemNoStop [0]{.\EOS\space}%
\providecommand \EOS [0]{\spacefactor3000\relax}%
\providecommand \BibitemShut  [1]{\csname bibitem#1\endcsname}%
\let\auto@bib@innerbib\@empty
\bibitem [{\citenamefont {Ginzburg}\ and\ \citenamefont
  {Landau}(1950)}]{Ginzburg:1950sr}%
  \BibitemOpen
  \bibfield  {author} {\bibinfo {author} {\bibfnamefont {V.~L.}\ \bibnamefont
  {Ginzburg}}\ and\ \bibinfo {author} {\bibfnamefont {L.~D.}\ \bibnamefont
  {Landau}},\ }\bibfield  {title} {\enquote {\bibinfo {title} {{On the Theory
  of superconductivity}},}\ }\href@noop {} {\bibfield  {journal} {\bibinfo
  {journal} {Zh. Eksp. Teor. Fiz.}\ }\textbf {\bibinfo {volume} {20}},\
  \bibinfo {pages} {1064--1082} (\bibinfo {year} {1950})}\BibitemShut {NoStop}%
\bibitem [{\citenamefont {Herbut}(2007)}]{herbut_2007}%
  \BibitemOpen
  \bibfield  {author} {\bibinfo {author} {\bibfnamefont {Igor}\ \bibnamefont
  {Herbut}},\ }\href {\doibase 10.1017/CBO9780511755521} {\emph {\bibinfo
  {title} {A Modern Approach to Critical Phenomena}}}\ (\bibinfo  {publisher}
  {Cambridge University Press},\ \bibinfo {year} {2007})\BibitemShut {NoStop}%
\bibitem [{\citenamefont {Kolnberger}\ and\ \citenamefont {Folk}(1990)}]{Folk}%
  \BibitemOpen
  \bibfield  {author} {\bibinfo {author} {\bibfnamefont {S.}~\bibnamefont
  {Kolnberger}}\ and\ \bibinfo {author} {\bibfnamefont {R.}~\bibnamefont
  {Folk}},\ }\bibfield  {title} {\enquote {\bibinfo {title} {Critical
  fluctuations in superconductors},}\ }\href {\doibase
  10.1103/PhysRevB.41.4083} {\bibfield  {journal} {\bibinfo  {journal} {Phys.
  Rev. B}\ }\textbf {\bibinfo {volume} {41}},\ \bibinfo {pages} {4083--4088}
  (\bibinfo {year} {1990})}\BibitemShut {NoStop}%
\bibitem [{\citenamefont {Dasgupta}\ and\ \citenamefont
  {Halperin}(1981)}]{Dasgupta:1981zz}%
  \BibitemOpen
  \bibfield  {author} {\bibinfo {author} {\bibfnamefont {C.}~\bibnamefont
  {Dasgupta}}\ and\ \bibinfo {author} {\bibfnamefont {B.~I.}\ \bibnamefont
  {Halperin}},\ }\bibfield  {title} {\enquote {\bibinfo {title} {{Phase
  Transition in a Lattice Model of Superconductivity}},}\ }\href {\doibase
  10.1103/PhysRevLett.47.1556} {\bibfield  {journal} {\bibinfo  {journal}
  {Phys. Rev. Lett.}\ }\textbf {\bibinfo {volume} {47}},\ \bibinfo {pages}
  {1556--1560} (\bibinfo {year} {1981})}\BibitemShut {NoStop}%
\bibitem [{\citenamefont {Halperin}\ \emph {et~al.}(1974)\citenamefont
  {Halperin}, \citenamefont {Lubensky},\ and\ \citenamefont
  {Ma}}]{Halperin:1973jh}%
  \BibitemOpen
  \bibfield  {author} {\bibinfo {author} {\bibfnamefont {B.~i.}\ \bibnamefont
  {Halperin}}, \bibinfo {author} {\bibfnamefont {T.~C.}\ \bibnamefont
  {Lubensky}}, \ and\ \bibinfo {author} {\bibfnamefont {Shang-keng}\
  \bibnamefont {Ma}},\ }\bibfield  {title} {\enquote {\bibinfo {title} {{First
  order phase transitions in superconductors and smectic A liquid crystals}},}\
  }\href {\doibase 10.1103/PhysRevLett.32.292} {\bibfield  {journal} {\bibinfo
  {journal} {Phys. Rev. Lett.}\ }\textbf {\bibinfo {volume} {32}},\ \bibinfo
  {pages} {292--295} (\bibinfo {year} {1974})}\BibitemShut {NoStop}%
\bibitem [{\citenamefont {Hindmarsh}\ \emph {et~al.}(2009)\citenamefont
  {Hindmarsh}, \citenamefont {Stuckey},\ and\ \citenamefont
  {Bevis}}]{Hindmarsh:2008dw}%
  \BibitemOpen
  \bibfield  {author} {\bibinfo {author} {\bibfnamefont {Mark}\ \bibnamefont
  {Hindmarsh}}, \bibinfo {author} {\bibfnamefont {Stephanie}\ \bibnamefont
  {Stuckey}}, \ and\ \bibinfo {author} {\bibfnamefont {Neil}\ \bibnamefont
  {Bevis}},\ }\bibfield  {title} {\enquote {\bibinfo {title} {{Abelian Higgs
  Cosmic Strings: Small Scale Structure and Loops}},}\ }\href {\doibase
  10.1103/PhysRevD.79.123504} {\bibfield  {journal} {\bibinfo  {journal} {Phys.
  Rev. D}\ }\textbf {\bibinfo {volume} {79}},\ \bibinfo {pages} {123504}
  (\bibinfo {year} {2009})},\ \Eprint {http://arxiv.org/abs/0812.1929}
  {arXiv:0812.1929 [hep-th]} \BibitemShut {NoStop}%
\bibitem [{\citenamefont {Kleinert}(1989)}]{Kleinert:1989kx}%
  \BibitemOpen
  \bibfield  {author} {\bibinfo {author} {\bibfnamefont {H.}~\bibnamefont
  {Kleinert}},\ }\href {\doibase 10.1142/0356} {\emph {\bibinfo {title} {Gauge
  Fields in Condensed Matter}}}\ (\bibinfo  {publisher} {WORLD SCIENTIFIC},\
  \bibinfo {year} {1989})\BibitemShut {NoStop}%
\bibitem [{\citenamefont {Elitzur}(1975)}]{Elitzur:1975im}%
  \BibitemOpen
  \bibfield  {author} {\bibinfo {author} {\bibfnamefont {S.}~\bibnamefont
  {Elitzur}},\ }\bibfield  {title} {\enquote {\bibinfo {title} {{Impossibility
  of Spontaneously Breaking Local Symmetries}},}\ }\href {\doibase
  10.1103/PhysRevD.12.3978} {\bibfield  {journal} {\bibinfo  {journal} {Phys.
  Rev. D}\ }\textbf {\bibinfo {volume} {12}},\ \bibinfo {pages} {3978--3982}
  (\bibinfo {year} {1975})}\BibitemShut {NoStop}%
\bibitem [{\citenamefont {Dirac}(1981)}]{dirac1981principles}%
  \BibitemOpen
  \bibfield  {author} {\bibinfo {author} {\bibfnamefont {Paul Adrien~Maurice}\
  \bibnamefont {Dirac}},\ }\href@noop {} {\emph {\bibinfo {title} {The
  principles of quantum mechanics}}},\ \bibinfo {number} {27}\ (\bibinfo
  {publisher} {Oxford university press},\ \bibinfo {year} {1981})\BibitemShut
  {NoStop}%
\bibitem [{\citenamefont {Schwinger}(1962)}]{Schwinger:1962tp}%
  \BibitemOpen
  \bibfield  {author} {\bibinfo {author} {\bibfnamefont {Julian~S.}\
  \bibnamefont {Schwinger}},\ }\bibfield  {title} {\enquote {\bibinfo {title}
  {{Gauge Invariance and Mass. 2.}}}\ }\href {\doibase
  10.1103/PhysRev.128.2425} {\bibfield  {journal} {\bibinfo  {journal} {Phys.
  Rev.}\ }\textbf {\bibinfo {volume} {128}},\ \bibinfo {pages} {2425--2429}
  (\bibinfo {year} {1962})}\BibitemShut {NoStop}%
\bibitem [{\citenamefont {Schwinger}(1959)}]{Schwinger:1959xd}%
  \BibitemOpen
  \bibfield  {author} {\bibinfo {author} {\bibfnamefont {Julian~S.}\
  \bibnamefont {Schwinger}},\ }\bibfield  {title} {\enquote {\bibinfo {title}
  {{Field theory commutators}},}\ }\href {\doibase 10.1103/PhysRevLett.3.296}
  {\bibfield  {journal} {\bibinfo  {journal} {Phys. Rev. Lett.}\ }\textbf
  {\bibinfo {volume} {3}},\ \bibinfo {pages} {296--297} (\bibinfo {year}
  {1959})}\BibitemShut {NoStop}%
\bibitem [{\citenamefont {Kennedy}\ and\ \citenamefont
  {King}(1985)}]{Kennedy:1985yn}%
  \BibitemOpen
  \bibfield  {author} {\bibinfo {author} {\bibfnamefont {Tom}\ \bibnamefont
  {Kennedy}}\ and\ \bibinfo {author} {\bibfnamefont {Chris}\ \bibnamefont
  {King}},\ }\bibfield  {title} {\enquote {\bibinfo {title} {{Symmetry Breaking
  in the Lattice Abelian Higgs Model}},}\ }\href {\doibase
  10.1103/PhysRevLett.55.776} {\bibfield  {journal} {\bibinfo  {journal} {Phys.
  Rev. Lett.}\ }\textbf {\bibinfo {volume} {55}},\ \bibinfo {pages} {776}
  (\bibinfo {year} {1985})}\BibitemShut {NoStop}%
\bibitem [{\citenamefont {Kleinert}\ and\ \citenamefont
  {Schakel}(2003)}]{Kleinert:2003}%
  \BibitemOpen
  \bibfield  {author} {\bibinfo {author} {\bibfnamefont {H.}~\bibnamefont
  {Kleinert}}\ and\ \bibinfo {author} {\bibfnamefont {Adriaan M.~J.}\
  \bibnamefont {Schakel}},\ }\bibfield  {title} {\enquote {\bibinfo {title}
  {{Gauge-Invariant Critical Exponents for the Ginzburg-Landau Model}},}\
  }\href {\doibase 10.1103/PhysRevLett.90.097001} {\bibfield  {journal}
  {\bibinfo  {journal} {Phys. Rev. Lett.}\ }\textbf {\bibinfo {volume} {90}},\
  \bibinfo {pages} {097001} (\bibinfo {year} {2003})}\BibitemShut {NoStop}%
\bibitem [{\citenamefont {Kleinert}\ and\ \citenamefont
  {Schakel}(2005)}]{Kleinert:2005sa}%
  \BibitemOpen
  \bibfield  {author} {\bibinfo {author} {\bibfnamefont {H.}~\bibnamefont
  {Kleinert}}\ and\ \bibinfo {author} {\bibfnamefont {Adriaan M.~J.}\
  \bibnamefont {Schakel}},\ }\bibfield  {title} {\enquote {\bibinfo {title}
  {{Anomalous dimension of Dirac's gauge-invariant nonlocal order parameter in
  Ginzburg-Landau field theory}},}\ }\href {\doibase
  10.1016/j.physletb.2005.02.044} {\bibfield  {journal} {\bibinfo  {journal}
  {Phys. Lett. B}\ }\textbf {\bibinfo {volume} {611}},\ \bibinfo {pages}
  {182--188} (\bibinfo {year} {2005})},\ \Eprint
  {http://arxiv.org/abs/cond-mat/0501036} {arXiv:cond-mat/0501036} \BibitemShut
  {NoStop}%
\bibitem [{\citenamefont {Frohlich}\ \emph {et~al.}(1981)\citenamefont
  {Frohlich}, \citenamefont {Morchio},\ and\ \citenamefont
  {Strocchi}}]{Frohlich:1981yi}%
  \BibitemOpen
  \bibfield  {author} {\bibinfo {author} {\bibfnamefont {J.}~\bibnamefont
  {Frohlich}}, \bibinfo {author} {\bibfnamefont {G.}~\bibnamefont {Morchio}}, \
  and\ \bibinfo {author} {\bibfnamefont {F.}~\bibnamefont {Strocchi}},\
  }\bibfield  {title} {\enquote {\bibinfo {title} {{HIGGS PHENOMENON WITHOUT
  SYMMETRY BREAKING ORDER PARAMETER}},}\ }\href {\doibase
  10.1016/0550-3213(81)90448-X} {\bibfield  {journal} {\bibinfo  {journal}
  {Nucl. Phys. B}\ }\textbf {\bibinfo {volume} {190}},\ \bibinfo {pages}
  {553--582} (\bibinfo {year} {1981})}\BibitemShut {NoStop}%
\bibitem [{\citenamefont {Hellerman}\ \emph {et~al.}(2015)\citenamefont
  {Hellerman}, \citenamefont {Orlando}, \citenamefont {Reffert},\ and\
  \citenamefont {Watanabe}}]{Hellerman:2015nra}%
  \BibitemOpen
  \bibfield  {author} {\bibinfo {author} {\bibfnamefont {Simeon}\ \bibnamefont
  {Hellerman}}, \bibinfo {author} {\bibfnamefont {Domenico}\ \bibnamefont
  {Orlando}}, \bibinfo {author} {\bibfnamefont {Susanne}\ \bibnamefont
  {Reffert}}, \ and\ \bibinfo {author} {\bibfnamefont {Masataka}\ \bibnamefont
  {Watanabe}},\ }\bibfield  {title} {\enquote {\bibinfo {title} {{On the CFT
  Operator Spectrum at Large Global Charge}},}\ }\href {\doibase
  10.1007/JHEP12(2015)071} {\bibfield  {journal} {\bibinfo  {journal} {JHEP}\
  }\textbf {\bibinfo {volume} {12}},\ \bibinfo {pages} {071} (\bibinfo {year}
  {2015})},\ \Eprint {http://arxiv.org/abs/1505.01537} {arXiv:1505.01537
  [hep-th]} \BibitemShut {NoStop}%
\bibitem [{\citenamefont {Gaum\'e}\ \emph {et~al.}(2021)\citenamefont
  {Gaum\'e}, \citenamefont {Orlando},\ and\ \citenamefont
  {Reffert}}]{Gaume:2020bmp}%
  \BibitemOpen
  \bibfield  {author} {\bibinfo {author} {\bibfnamefont {Luis~\'Alvarez}\
  \bibnamefont {Gaum\'e}}, \bibinfo {author} {\bibfnamefont {Domenico}\
  \bibnamefont {Orlando}}, \ and\ \bibinfo {author} {\bibfnamefont {Susanne}\
  \bibnamefont {Reffert}},\ }\bibfield  {title} {\enquote {\bibinfo {title}
  {{Selected topics in the large quantum number expansion}},}\ }\href {\doibase
  10.1016/j.physrep.2021.08.001} {\bibfield  {journal} {\bibinfo  {journal}
  {Phys. Rept.}\ }\textbf {\bibinfo {volume} {933}},\ \bibinfo {pages} {1--66}
  (\bibinfo {year} {2021})},\ \Eprint {http://arxiv.org/abs/2008.03308}
  {arXiv:2008.03308 [hep-th]} \BibitemShut {NoStop}%
\bibitem [{\citenamefont {Badel}\ \emph {et~al.}(2019)\citenamefont {Badel},
  \citenamefont {Cuomo}, \citenamefont {Monin},\ and\ \citenamefont
  {Rattazzi}}]{Badel:2019oxl}%
  \BibitemOpen
  \bibfield  {author} {\bibinfo {author} {\bibfnamefont {Gil}\ \bibnamefont
  {Badel}}, \bibinfo {author} {\bibfnamefont {Gabriel}\ \bibnamefont {Cuomo}},
  \bibinfo {author} {\bibfnamefont {Alexander}\ \bibnamefont {Monin}}, \ and\
  \bibinfo {author} {\bibfnamefont {Riccardo}\ \bibnamefont {Rattazzi}},\
  }\bibfield  {title} {\enquote {\bibinfo {title} {{The Epsilon Expansion Meets
  Semiclassics}},}\ }\href {\doibase 10.1007/JHEP11(2019)110} {\bibfield
  {journal} {\bibinfo  {journal} {JHEP}\ }\textbf {\bibinfo {volume} {11}},\
  \bibinfo {pages} {110} (\bibinfo {year} {2019})},\ \Eprint
  {http://arxiv.org/abs/1909.01269} {arXiv:1909.01269 [hep-th]} \BibitemShut
  {NoStop}%
\bibitem [{\citenamefont {Antipin}\ \emph
  {et~al.}(2020{\natexlab{a}})\citenamefont {Antipin}, \citenamefont {Bersini},
  \citenamefont {Sannino}, \citenamefont {Wang},\ and\ \citenamefont
  {Zhang}}]{Antipin:2020abu}%
  \BibitemOpen
  \bibfield  {author} {\bibinfo {author} {\bibfnamefont {Oleg}\ \bibnamefont
  {Antipin}}, \bibinfo {author} {\bibfnamefont {Jahmall}\ \bibnamefont
  {Bersini}}, \bibinfo {author} {\bibfnamefont {Francesco}\ \bibnamefont
  {Sannino}}, \bibinfo {author} {\bibfnamefont {Zhi-Wei}\ \bibnamefont {Wang}},
  \ and\ \bibinfo {author} {\bibfnamefont {Chen}\ \bibnamefont {Zhang}},\
  }\bibfield  {title} {\enquote {\bibinfo {title} {{Charging the $O(N)$
  model}},}\ }\href {\doibase 10.1103/PhysRevD.102.045011} {\bibfield
  {journal} {\bibinfo  {journal} {Phys. Rev. D}\ }\textbf {\bibinfo {volume}
  {102}},\ \bibinfo {pages} {045011} (\bibinfo {year} {2020}{\natexlab{a}})},\
  \Eprint {http://arxiv.org/abs/2003.13121} {arXiv:2003.13121 [hep-th]}
  \BibitemShut {NoStop}%
\bibitem [{\citenamefont {Monin}\ \emph {et~al.}(2017)\citenamefont {Monin},
  \citenamefont {Pirtskhalava}, \citenamefont {Rattazzi},\ and\ \citenamefont
  {Seibold}}]{Monin:2016jmo}%
  \BibitemOpen
  \bibfield  {author} {\bibinfo {author} {\bibfnamefont {Alexander}\
  \bibnamefont {Monin}}, \bibinfo {author} {\bibfnamefont {David}\ \bibnamefont
  {Pirtskhalava}}, \bibinfo {author} {\bibfnamefont {Riccardo}\ \bibnamefont
  {Rattazzi}}, \ and\ \bibinfo {author} {\bibfnamefont {Fiona~K.}\ \bibnamefont
  {Seibold}},\ }\bibfield  {title} {\enquote {\bibinfo {title} {{Semiclassics,
  Goldstone Bosons and CFT data}},}\ }\href {\doibase 10.1007/JHEP06(2017)011}
  {\bibfield  {journal} {\bibinfo  {journal} {JHEP}\ }\textbf {\bibinfo
  {volume} {06}},\ \bibinfo {pages} {011} (\bibinfo {year} {2017})},\ \Eprint
  {http://arxiv.org/abs/1611.02912} {arXiv:1611.02912 [hep-th]} \BibitemShut
  {NoStop}%
\bibitem [{\citenamefont {Alvarez-Gaume}\ \emph {et~al.}(2017)\citenamefont
  {Alvarez-Gaume}, \citenamefont {Loukas}, \citenamefont {Orlando},\ and\
  \citenamefont {Reffert}}]{Alvarez-Gaume:2016vff}%
  \BibitemOpen
  \bibfield  {author} {\bibinfo {author} {\bibfnamefont {Luis}\ \bibnamefont
  {Alvarez-Gaume}}, \bibinfo {author} {\bibfnamefont {Orestis}\ \bibnamefont
  {Loukas}}, \bibinfo {author} {\bibfnamefont {Domenico}\ \bibnamefont
  {Orlando}}, \ and\ \bibinfo {author} {\bibfnamefont {Susanne}\ \bibnamefont
  {Reffert}},\ }\bibfield  {title} {\enquote {\bibinfo {title} {{Compensating
  strong coupling with large charge}},}\ }\href {\doibase
  10.1007/JHEP04(2017)059} {\bibfield  {journal} {\bibinfo  {journal} {JHEP}\
  }\textbf {\bibinfo {volume} {04}},\ \bibinfo {pages} {059} (\bibinfo {year}
  {2017})},\ \Eprint {http://arxiv.org/abs/1610.04495} {arXiv:1610.04495
  [hep-th]} \BibitemShut {NoStop}%
\bibitem [{\citenamefont {Banerjee}\ \emph {et~al.}(2018)\citenamefont
  {Banerjee}, \citenamefont {Chandrasekharan},\ and\ \citenamefont
  {Orlando}}]{Banerjee:2017fcx}%
  \BibitemOpen
  \bibfield  {author} {\bibinfo {author} {\bibfnamefont {Debasish}\
  \bibnamefont {Banerjee}}, \bibinfo {author} {\bibfnamefont {Shailesh}\
  \bibnamefont {Chandrasekharan}}, \ and\ \bibinfo {author} {\bibfnamefont
  {Domenico}\ \bibnamefont {Orlando}},\ }\bibfield  {title} {\enquote {\bibinfo
  {title} {{Conformal dimensions via large charge expansion}},}\ }\href
  {\doibase 10.1103/PhysRevLett.120.061603} {\bibfield  {journal} {\bibinfo
  {journal} {Phys. Rev. Lett.}\ }\textbf {\bibinfo {volume} {120}},\ \bibinfo
  {pages} {061603} (\bibinfo {year} {2018})},\ \Eprint
  {http://arxiv.org/abs/1707.00711} {arXiv:1707.00711 [hep-lat]} \BibitemShut
  {NoStop}%
\bibitem [{\citenamefont {Giombi}\ \emph {et~al.}(2022)\citenamefont {Giombi},
  \citenamefont {Helfenberger},\ and\ \citenamefont
  {Khanchandani}}]{Giombi:2022gjj}%
  \BibitemOpen
  \bibfield  {author} {\bibinfo {author} {\bibfnamefont {Simone}\ \bibnamefont
  {Giombi}}, \bibinfo {author} {\bibfnamefont {Elizabeth}\ \bibnamefont
  {Helfenberger}}, \ and\ \bibinfo {author} {\bibfnamefont {Himanshu}\
  \bibnamefont {Khanchandani}},\ }\bibfield  {title} {\enquote {\bibinfo
  {title} {{Long Range, Large Charge, Large $N$}},}\ }\href@noop {} {\
  (\bibinfo {year} {2022})},\ \Eprint {http://arxiv.org/abs/2205.00500}
  {arXiv:2205.00500 [hep-th]} \BibitemShut {NoStop}%
\bibitem [{\citenamefont {Antipin}\ \emph {et~al.}(2022)\citenamefont
  {Antipin}, \citenamefont {Bersini},\ and\ \citenamefont
  {Panopoulos}}]{Antipin:2022naw}%
  \BibitemOpen
  \bibfield  {author} {\bibinfo {author} {\bibfnamefont {Oleg}\ \bibnamefont
  {Antipin}}, \bibinfo {author} {\bibfnamefont {Jahmall}\ \bibnamefont
  {Bersini}}, \ and\ \bibinfo {author} {\bibfnamefont {Pantelis}\ \bibnamefont
  {Panopoulos}},\ }\bibfield  {title} {\enquote {\bibinfo {title} {{Yukawa
  interactions at large charge}},}\ }\href {\doibase 10.1007/JHEP10(2022)183}
  {\bibfield  {journal} {\bibinfo  {journal} {JHEP}\ }\textbf {\bibinfo
  {volume} {10}},\ \bibinfo {pages} {183} (\bibinfo {year} {2022})},\ \Eprint
  {http://arxiv.org/abs/2208.05839} {arXiv:2208.05839 [hep-th]} \BibitemShut
  {NoStop}%
\bibitem [{\citenamefont {Cardy}(1984)}]{Cardy:1984rp}%
  \BibitemOpen
  \bibfield  {author} {\bibinfo {author} {\bibfnamefont {J.~L.}\ \bibnamefont
  {Cardy}},\ }\bibfield  {title} {\enquote {\bibinfo {title} {{Conformal
  invariance and universality in finite-size scaling}},}\ }\href@noop {}
  {\bibfield  {journal} {\bibinfo  {journal} {J. Phys. A}\ }\textbf {\bibinfo
  {volume} {17}},\ \bibinfo {pages} {L385--L387} (\bibinfo {year}
  {1984})}\BibitemShut {NoStop}%
\bibitem [{\citenamefont {Cardy}(1985)}]{Cardy:1985lth}%
  \BibitemOpen
  \bibfield  {author} {\bibinfo {author} {\bibfnamefont {J.~L.}\ \bibnamefont
  {Cardy}},\ }\bibfield  {title} {\enquote {\bibinfo {title} {{Universal
  amplitudes in finite-size scaling: generalisation to arbitrary
  dimensionality}},}\ }\href {\doibase 10.1088/0305-4470/18/13/005} {\bibfield
  {journal} {\bibinfo  {journal} {J. Phys. A}\ }\textbf {\bibinfo {volume}
  {18}},\ \bibinfo {pages} {L757--L760} (\bibinfo {year} {1985})}\BibitemShut
  {NoStop}%
\bibitem [{\citenamefont {Coleman}\ and\ \citenamefont
  {Weinberg}(1973)}]{Coleman:1973jx}%
  \BibitemOpen
  \bibfield  {author} {\bibinfo {author} {\bibfnamefont {Sidney~R.}\
  \bibnamefont {Coleman}}\ and\ \bibinfo {author} {\bibfnamefont {Erick~J.}\
  \bibnamefont {Weinberg}},\ }\bibfield  {title} {\enquote {\bibinfo {title}
  {{Radiative Corrections as the Origin of Spontaneous Symmetry Breaking}},}\
  }\href {\doibase 10.1103/PhysRevD.7.1888} {\bibfield  {journal} {\bibinfo
  {journal} {Phys. Rev. D}\ }\textbf {\bibinfo {volume} {7}},\ \bibinfo {pages}
  {1888--1910} (\bibinfo {year} {1973})}\BibitemShut {NoStop}%
\bibitem [{\citenamefont {Antipin}\ \emph
  {et~al.}(2020{\natexlab{b}})\citenamefont {Antipin}, \citenamefont {Bersini},
  \citenamefont {Sannino}, \citenamefont {Wang},\ and\ \citenamefont
  {Zhang}}]{Antipin:2020rdw}%
  \BibitemOpen
  \bibfield  {author} {\bibinfo {author} {\bibfnamefont {Oleg}\ \bibnamefont
  {Antipin}}, \bibinfo {author} {\bibfnamefont {Jahmall}\ \bibnamefont
  {Bersini}}, \bibinfo {author} {\bibfnamefont {Francesco}\ \bibnamefont
  {Sannino}}, \bibinfo {author} {\bibfnamefont {Zhi-Wei}\ \bibnamefont {Wang}},
  \ and\ \bibinfo {author} {\bibfnamefont {Chen}\ \bibnamefont {Zhang}},\
  }\bibfield  {title} {\enquote {\bibinfo {title} {{Charging non-Abelian Higgs
  theories}},}\ }\href {\doibase 10.1103/PhysRevD.102.125033} {\bibfield
  {journal} {\bibinfo  {journal} {Phys. Rev. D}\ }\textbf {\bibinfo {volume}
  {102}},\ \bibinfo {pages} {125033} (\bibinfo {year} {2020}{\natexlab{b}})},\
  \Eprint {http://arxiv.org/abs/2006.10078} {arXiv:2006.10078 [hep-th]}
  \BibitemShut {NoStop}%
\bibitem [{\citenamefont {Brown}\ and\ \citenamefont
  {Collins}(1980)}]{Brown:1980qq}%
  \BibitemOpen
  \bibfield  {author} {\bibinfo {author} {\bibfnamefont {Lowell~S.}\
  \bibnamefont {Brown}}\ and\ \bibinfo {author} {\bibfnamefont {John~C.}\
  \bibnamefont {Collins}},\ }\bibfield  {title} {\enquote {\bibinfo {title}
  {{Dimensional Renormalization of Scalar Field Theory in Curved
  Space-time}},}\ }\href {\doibase 10.1016/0003-4916(80)90232-8} {\bibfield
  {journal} {\bibinfo  {journal} {Annals Phys.}\ }\textbf {\bibinfo {volume}
  {130}},\ \bibinfo {pages} {215} (\bibinfo {year} {1980})}\BibitemShut
  {NoStop}%
\bibitem [{\citenamefont {Vladimirov}(1980)}]{Vladimirov:1979zm}%
  \BibitemOpen
  \bibfield  {author} {\bibinfo {author} {\bibfnamefont {A.~A.}\ \bibnamefont
  {Vladimirov}},\ }\bibfield  {title} {\enquote {\bibinfo {title} {{Method for
  Computing Renormalization Group Functions in Dimensional Renormalization
  Scheme}},}\ }\href {\doibase 10.1007/BF01018394} {\bibfield  {journal}
  {\bibinfo  {journal} {Theor. Math. Phys.}\ }\textbf {\bibinfo {volume}
  {43}},\ \bibinfo {pages} {417} (\bibinfo {year} {1980})}\BibitemShut
  {NoStop}%
\bibitem [{\citenamefont {Steinhauser}(2001)}]{Steinhauser:2000ry}%
  \BibitemOpen
  \bibfield  {author} {\bibinfo {author} {\bibfnamefont {Matthias}\
  \bibnamefont {Steinhauser}},\ }\bibfield  {title} {\enquote {\bibinfo {title}
  {{MATAD: A Program package for the computation of MAssive TADpoles}},}\
  }\href {\doibase 10.1016/S0010-4655(00)00204-6} {\bibfield  {journal}
  {\bibinfo  {journal} {Comput. Phys. Commun.}\ }\textbf {\bibinfo {volume}
  {134}},\ \bibinfo {pages} {335--364} (\bibinfo {year} {2001})},\ \Eprint
  {http://arxiv.org/abs/hep-ph/0009029} {arXiv:hep-ph/0009029} \BibitemShut
  {NoStop}%
\bibitem [{\citenamefont {de~Calan}\ and\ \citenamefont
  {Nogueira}(1999)}]{deCalan:1999vx}%
  \BibitemOpen
  \bibfield  {author} {\bibinfo {author} {\bibfnamefont {Claude}\ \bibnamefont
  {de~Calan}}\ and\ \bibinfo {author} {\bibfnamefont {Flavio~S.}\ \bibnamefont
  {Nogueira}},\ }\bibfield  {title} {\enquote {\bibinfo {title} {{The
  Superconducting order parameter and gauge dependence}},}\ }\href@noop {} {\
  (\bibinfo {year} {1999})},\ \Eprint {http://arxiv.org/abs/cond-mat/9905276}
  {arXiv:cond-mat/9905276} \BibitemShut {NoStop}%
\bibitem [{\citenamefont {Jack}\ and\ \citenamefont
  {Jones}(2021)}]{Jack:2021ypd}%
  \BibitemOpen
  \bibfield  {author} {\bibinfo {author} {\bibfnamefont {I.}~\bibnamefont
  {Jack}}\ and\ \bibinfo {author} {\bibfnamefont {D.~R.~T.}\ \bibnamefont
  {Jones}},\ }\bibfield  {title} {\enquote {\bibinfo {title} {{Anomalous
  dimensions at large charge in d=4 O(N) theory}},}\ }\href {\doibase
  10.1103/PhysRevD.103.085013} {\bibfield  {journal} {\bibinfo  {journal}
  {Phys. Rev. D}\ }\textbf {\bibinfo {volume} {103}},\ \bibinfo {pages}
  {085013} (\bibinfo {year} {2021})},\ \Eprint
  {http://arxiv.org/abs/2101.09820} {arXiv:2101.09820 [hep-th]} \BibitemShut
  {NoStop}%
\bibitem [{\citenamefont {Bednyakov}\ and\ \citenamefont
  {Pikelner}(2022)}]{Bednyakov:2022guj}%
  \BibitemOpen
  \bibfield  {author} {\bibinfo {author} {\bibfnamefont {Alexander}\
  \bibnamefont {Bednyakov}}\ and\ \bibinfo {author} {\bibfnamefont {Andrey}\
  \bibnamefont {Pikelner}},\ }\bibfield  {title} {\enquote {\bibinfo {title}
  {{Six-loop anomalous dimension of the $\ensuremath{\phi^Q}$ operator in the
  $O(N)$ symmetric model}},}\ }\href {\doibase 10.1103/PhysRevD.106.076015}
  {\bibfield  {journal} {\bibinfo  {journal} {Phys. Rev. D}\ }\textbf {\bibinfo
  {volume} {106}},\ \bibinfo {pages} {076015} (\bibinfo {year} {2022})},\
  \Eprint {http://arxiv.org/abs/2208.04612} {arXiv:2208.04612 [hep-th]}
  \BibitemShut {NoStop}%
\end{thebibliography}%
\end{document}